\documentclass[prl,amsmath,twocolumn,amssymb,aps,superscriptaddress,floatfix,preprintnumbers,notitlepage,10pt]{revtex4-1}
\usepackage{times}
\usepackage{graphicx}
\usepackage{amsmath, braket, amsfonts}
\usepackage{amssymb}
\usepackage{bm, color, ulem}

\definecolor{pink}{rgb}{0.858, 0.188, 0.478}

\def\lsim{\lower -0.3ex \hbox{$<$} \kern -0.75em \lower 0.7ex \hbox{$\sim$}}
\def\gsim{\lower -0.3ex \hbox{$>$} \kern -0.75em \lower 0.7ex \hbox{$\sim$}}
\def\mb{\bf}
\def\Vec#1{{\bf #1}}
\def\GVec#1{\mbox{\boldmath $#1$}}
\def\t{\theta}
\def\e{\epsilon}
\def\H{{\mathcal H}}
\def\vare{\varepsilon}
\def\av#1{\langle #1 \rangle}
\def\sgn{{\rm sg	n}}
\def\partd#1#2{\frac{\partial #1}{\partial #2}}
\def\partdd#1#2{\frac{\partial^2 #1}{\partial #2^2}}
\def\kappaval{3}

\newcommand{\figpath}{./figs_v15_lowres.pdf}

\begin{document}
\title{Even denominator fractional quantum Hall states at an isospin transition in monolayer graphene}

\author{A.A. Zibrov*}
\affiliation{Department of Physics, University of California, Santa Barbara CA 93106 USA}
\author{E.M. Spanton*}
\affiliation{California Nanosystems Institute, University of California at Santa Barbara, Santa Barbara, CA, 93106}
\author{H. Zhou}
\affiliation{Department of Physics, University of California, Santa Barbara CA 93106 USA}
\author{C. Kometter}
\affiliation{Department of Physics, University of California, Santa Barbara CA 93106 USA}
\author{T. Taniguchi}
\affiliation{Advanced Materials Laboratory, National Institute for Materials Science, Tsukuba, Ibaraki 305-0044, Japan}
\author{K. Watanabe}
\affiliation{Advanced Materials Laboratory, National Institute for Materials Science, Tsukuba, Ibaraki 305-0044, Japan}
\author{A.F. Young}
\affiliation{Department of Physics, University of California, Santa Barbara CA 93106 USA}
\date{\today}

\begin{abstract}
Magnetic fields quench the kinetic energy of two dimensional electrons, confining them to highly degenerate Landau levels.  In the absence of disorder, the ground state at partial Landau level filling is determined only by Coulomb interactions, leading to a variety of correlation-driven phenomena.
Here, we realize a quantum Hall analog of the Ne\'el-to-valence bond solid transition  within
the spin- and sublattice- degenerate monolayer graphene zero energy Landau level by experimentally controlling substrate-induced sublattice symmetry breaking. The transition is marked by unusual isospin transitions in odd denominator fractional quantum Hall states for filling factors $\nu$ near charge neutrality, and the unexpected appearance of incompressible even denominator fractional quantum Hall states at $\nu=\pm1/2$ and $\pm1/4$ associated with pairing between composite fermions on different carbon sublattices.
\end{abstract}
\maketitle

Clean two dimensional electron systems in the high magnetic field limit host various correlated phenomena including Wigner crystallization of electrons, topologically ordered fractional quantum Hall liquids, and quantum Hall ferromagnets. Among such systems, monolayer graphene is distinguished by its zero energy Landau level (ZLL), which is  characterized by a near degeneracy of four isospin flavors arising from the degeneracy between Landau level orbitals with differing spin and sublattice polarizations. The dominant long-ranged Coulomb interaction does not distinguish between different spin or sublattice flavors, but favors full polarization within the resulting approximately SU(4) symmetric isospin\cite{nomura_quantum_2006}.  The direction of polarization is instead set by competing isospin anisotropies, including both single particle effects and the anisotropy of the Coulomb interactions at the scale of the honeycomb lattice.  Proposed ground states at charge neutrality include states characterized by either spin or sublattice order, such as a canted antiferromagnetic (CAF)\cite{herbut_theory_2007} state that breaks spin symmetry and a partially sublattice polarized (PSP) density wave\cite{nomura_field-induced_2009} that triples the size of the unit cell.

The CAF and PSP states are direct analogs of the Ne\'el and valence bond solid (VBS) states that arise in studies of two dimensional quantum magnetism, as noted in a series of recent theoretical works\cite{lee_deconfined_2014, wu_so5_2014, lee_wess-zumino-witten_2015}.  Within conventional Landau-Ginzburg-Wilson theory, incompatible symmetry breaking between the VBS and Ne\'el phases (real-space and spin, respectively) require a first order transition. However, unusual critical phases allowing for a continuous transition have been proposed\cite{senthil_deconfined_2004}, as well as first order transitions with emergent symmetry at the critical point\cite{wu_so5_2014}.
Realizing the PSP-CAF transition in monolayer could allow direct experimental probes of this unconventional quantum phase transition.

Here we report the observation of an isospin transition in monolayer graphene associated with the PSP-CAF phase boundary, realized by balancing intrinsic antiferromagnetic interaction anisotropy\cite{young_tunable_2014} against a substrate-induced sublattice-symmetry breaking gap\cite{hunt_massive_2013,amet_insulating_2013}.
The transition is marked by the appearance---and subsequent disappearance---of even-denominator fractional quantum Hall (EDFQH) states at $\nu=\pm1/2$ and $\pm1/4$, coincident in magnetic field with weakening of nearby odd denominator fractional quantum Hall (ODFQH) states.  We observe similar phenomenology in three monolayer graphene samples (A, B, and C) fabricated by encapsulating the graphene flake between single crystal hexagonal boron nitride gate dielectrics and single crystal graphite electrostatic gates\cite{zibrov_tunable_2017,li_even_2017}.  Across all devices, the isospin transition magnetic field varies by a factor of 5, and is directly correlated with sublattice splitting $\Delta_{AB}$, as expected from a mean field analysis of the proximal charge neutral state\cite{kharitonov_phase_2012, kharitonov_canted_2012}.  However, as we describe below, the details of the ODFQH, and indeed the very existence of EDFQH states are not captured by currently available theory for fractional quantum Hall states in monolayer graphene\cite{apalkov_fractional_2006, toke_theoretical_2007, abanin_fractional_2010, papic_tunable_2011, peterson_effects_2014}.



\begin{figure*}[ht!]
\begin{center}
\includegraphics[width=183mm, page=1]{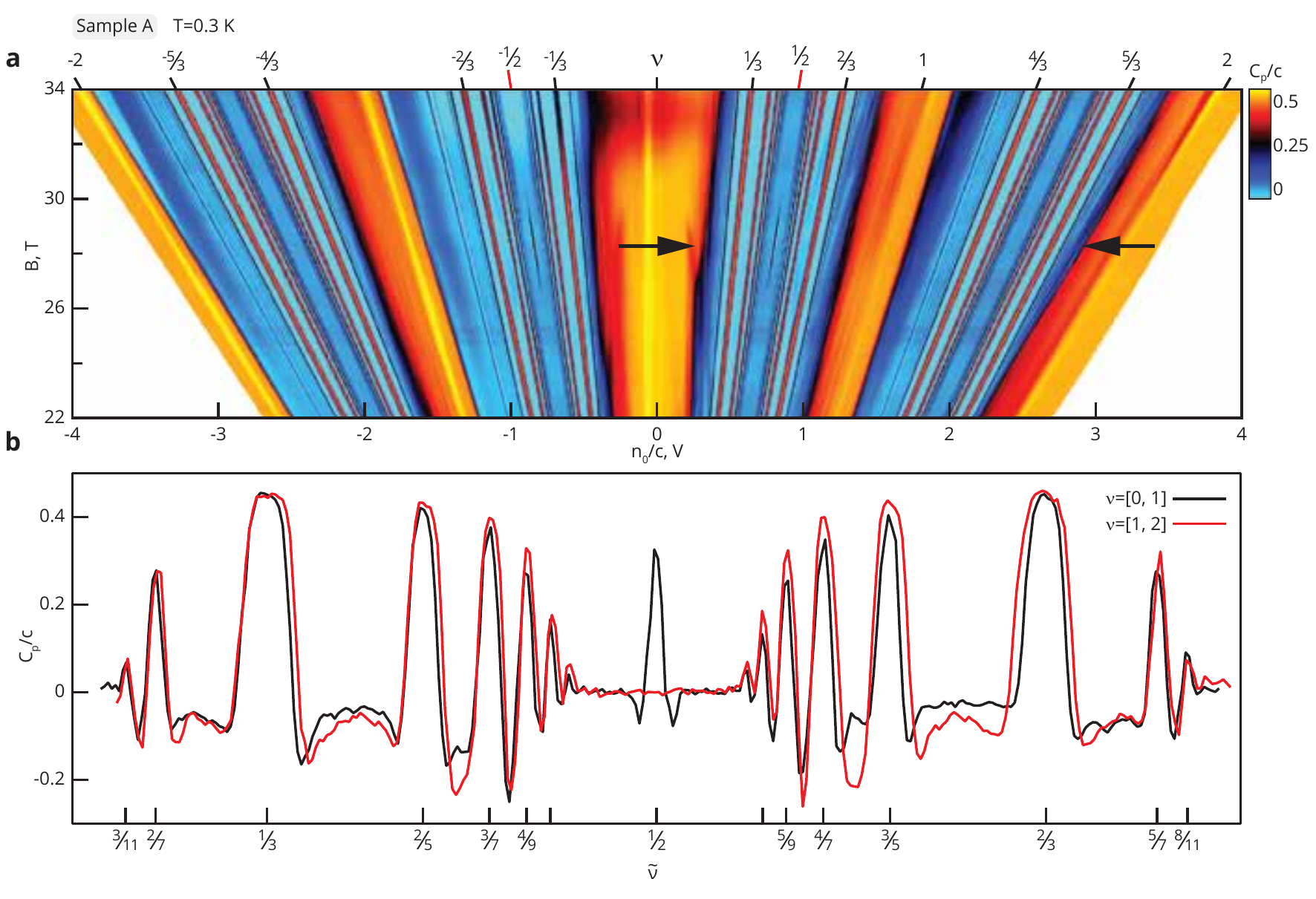}
\caption{ \textbf{Incompressible FQH states at $\mathbf{\nu = \pm 1/2}$.}
\textbf{(a)} False color plot of penetration field capacitance $C_\mathrm{P}/c$ in sample A as a function of the the magnetic field, $B$, and the nominal charge density $n_0 \equiv c(v_{t} + v_{b} - 2 v_{s})$ where $v_s$, $v_b$ and $v_s$ are the top gate, bottom gate, and sample voltages, respectively, and $c$ is the average geometric capacitance of the two gates (see Methods). Fractional quantum Hall states appear as lines of high $C_P$ with a slope proportional to their quantized Hall conductivity (see Methods).  The dataset spans filling factors $\nu=[-2, 2]$, encompassing the zero energy Landau level.
\textbf{(b)} Traces taken at constant $B=28.3$~T between filling factor $\nu=0$ and $\nu=2$ filling (indicated by black arrows in a).  $C_\mathrm{P}$ is  plotted as a function of relative filling factor $\tilde\nu\equiv\nu-\lfloor\nu\rfloor$, revealing a robust incompressible state
at half-filling of the $\nu=[0, 1]$ LL (black line) and a fully compressible state at half filling of the $\nu=[1, 2]$ LL (red line).
\label{fig1}
}
\end{center}
\end{figure*}

Figure \ref{fig1}a shows penetration field capacitance $C_\mathrm{P}$ for sample A, where $C_\mathrm{P}$ is defined as the differential capacitance between the top and bottom gates with the graphene held at constant electrical potential\cite{eisenstein_negative_1992}.  Data are plotted over a range spanning the ZLL as a function  of magnetic field $B$ and nominal charge carrier density $n_0=c(v_t+v_b - 2v_s)$, where $c$ is the average gate-to-sample geometric capacitance  of the two (nearly symmetric) gates and $v_t$, $v_b$, and $v_s$ are the voltages applied to top gate, bottom gate, and sample, respectively.
We observe gapped quantum Hall states, which appear as peaks in the measured signal (see Methods),
at integer fillings $\nu=\pm2, \pm1, 0$, as well as at filling factors associated with integer quantum Hall effects of both two and four-flux composite fermions\cite{jain_composite-fermion_1989}, $|\nu -\lfloor\nu\rfloor|= \frac{p}{mp\pm1}$, with $m=2,4$ and $p$ large as 7.

\begin{figure}[ht!]
\begin{center}
\includegraphics[width=89mm, page=2]{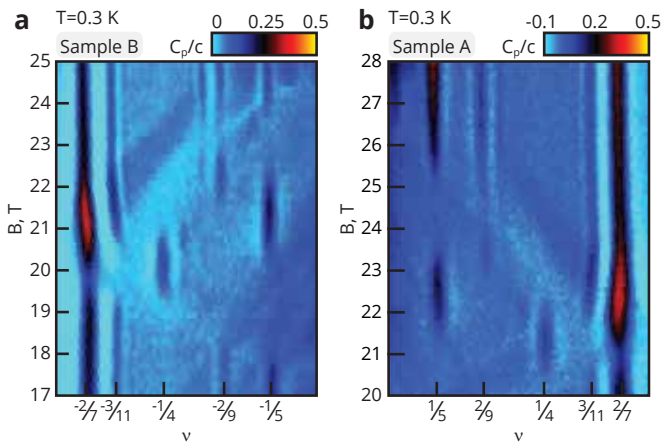}
\caption{
 \textbf{Incompressible FQH states at $\mathbf{\nu = \pm 1/4}$.}
\textbf{(a)} False color plot of $C_\mathrm{P}/c$ in sample B as a function of $\nu$ and $B$ at T=300 mK.  In addition to the $\nu=-1/4$ state, FQH states from the four-flux composite fermion sequence are visible at $\nu=\frac{p}{4p+1}$, with $B$-dependent weakening associated with isospin phase transitions.
\textbf{(b)} Similar data from sample A near $\nu=+1/4$.
\label{fourflux}
}
\end{center}
\end{figure}

Incompressible EDFQH states appear at $\nu = \pm 1/2$, but only in a narrow range of magnetic fields ($B=26.5$--$30$ T for sample A).  In contrast, $\nu=\pm3/2$ both remain compressible (Fig.~\ref{fig1}b). Incompressible EDFQH states are also observed at $\nu=\pm1/4$ in both Samples A and B, again appearing only for a small range of magnetic fields (Fig.~\ref{fourflux} and Fig.~S1). As with the $\pm1/2$ states, EDFQH is observed only near $\nu=0$, with no incompressible states observed at $\nu=\pm3/4$, $\pm5/4$ or $\pm7/4$.


\begin{figure*}[ht!]
\begin{center}
\includegraphics[width=183mm, page=3]{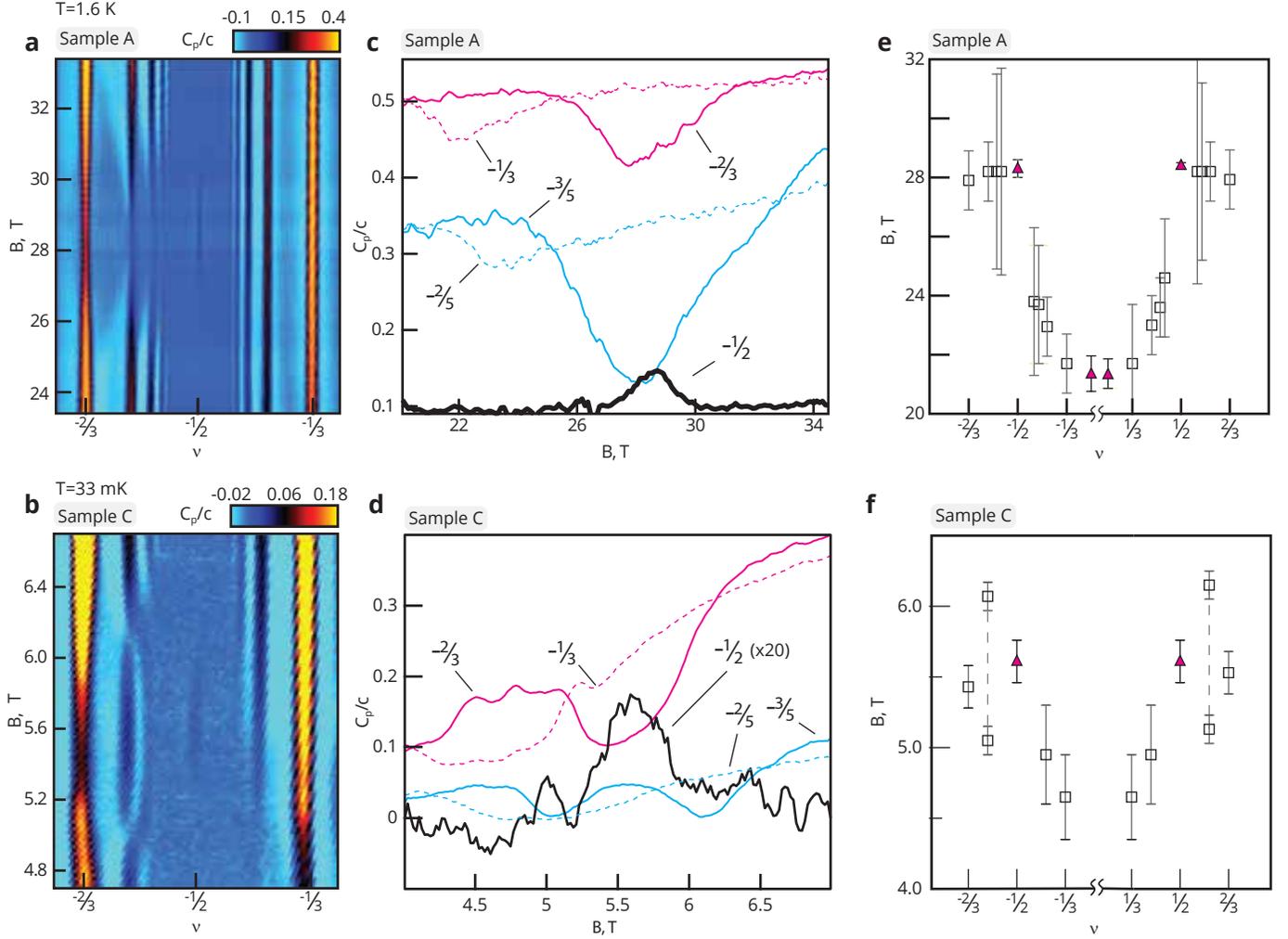}
\caption{
\textbf{Odd denominator fractional quantum Hall phase transitions associated with the $\mathbf{\nu = - 1/2}$ state.}
\textbf{(a)} $C_P$ as a function of $B$ and $\nu$ in the vicinity of $\nu=-1/2$ state for sample A, taken at $T=1.6$ K near B=28T.
\textbf{(b)}) Similar data from Sample C, taken at $T=33$ mK for B near 5.6T.
\textbf{(c,d)} $C_\mathrm{P}$ peak height as a function of magnetic field plotted for selected FQH states $\nu\in(-1, 0)$ for sample A \textbf{(c)} and
\textbf{(d)} sample C, showing the simultaneous strengthening of the even denominator state and weakening of adjacent odd-denominator states.
\textbf{(e)} Positions of minima (maxima) of the odd (even)-denominator FQH states for $\nu\in(-1, 1)$ in Sample A at T=1.6 K.
\textbf{(f)} Similar analysis of sample C at T= 33 mK. The density dependence of the transition field is not captured by our mean field model\cite{kharitonov_phase_2012,kharitonov_canted_2012}, which would imply that near $\nu=\pm1$, the phase transition should be at arbitrarily high magnetic field. In sample C, two gap closings are observed at $\nu = \pm 3/5$, while only one a single gap weakening is observed in sample A.
\label{ABcomparison}
}
\end{center}
\end{figure*}

The appearance of EDFQH states is accompanied by weakening or disappearance of the ODFQH states.  This is evident in the four-flux sequence (Fig.~\ref{fourflux}) as well as in the two-flux states near $\nu = \pm 1/2$, as shown in \ref{ABcomparison}a-d and Fig.~S2. The weakening is more evident at temperatures comparable to the ODFQH energy gaps, as seen in Fig.~S12.  The magnetic field at which we observe the ODFQH weakening or EDFQH emergence is density dependent in all samples, with weakening occurring at lower fields for transitions closer to $\nu = 0$ (Fig.~\ref{ABcomparison}e-f). No similar weakening is observed for any FQH states in $|\nu|\in(1,2)$, nor for four-flux ODFQH states near $\nu=\pm3/4$ (see Fig. 1a and supplementary Figs.~S3-5).


The $\nu$ dependence of the anomalous EDFQH and ODFQH phenomenology is consistent with an underlying isospin phase transition occurring at $\nu=0$.  The effect of the intrinsic anisotropy of the Coulomb interactions strengthens near charge neutrality, which corresponds to half filling of the ZLL. At charge neutrality, the phase diagram is dominated by the effects of these interactions, which also influence the nature of nearby FQH phases.  For example, interaction anisotropies increase the observed energy gaps for  $|\nu|<1$ FQH states as compared to $1<|\nu|<2$ states\cite{dean_multicomponent_2011, feldman_unconventional_2012}, where their absence permits the formation of large, low energy valley (equivalent to sublattice) skyrmions in some ODFQH states.  In addition, isospin phase transitions have previously been observed in monolayer graphene for $|\nu|<1$, with no corresponding phase transitions for $|\nu|>1$\cite{feldman_fractional_2013}.  These transitions were explained as arising from a Zeeman-effect driven crossing of composite fermion Landau levels associated with the isospin order at $\nu=0$\cite{abanin_fractional_2013,sodemann_broken_2014}. As with the data sets described here, transitions in Ref. \onlinecite{feldman_fractional_2013} occur at higher $B$ for higher $|\nu|$.

However, EDFQH has not been reported among the many FQH states observed in monolayer graphene\cite{bolotin_ultrahigh_2008, du_fractional_2009, dean_multicomponent_2011, feldman_unconventional_2012, feldman_fractional_2013,amet_composite_2015}, nor have EDFQH states been predicted\cite{apalkov_fractional_2006, toke_theoretical_2007, toke_su4_2007,shibata_fractional_2009, papic_atypical_2010, toke_multi-component_2011, papic_tunable_2011, peterson_effects_2014}.
Previous experiments on other quantum Hall systems have revealed contrasting behavior in half-filled Landau levels.
In single layer semiconductor quantum wells, the 2D electron system is compressible at filling factors $\nu=1/2, 3/2$ (corresponding to the lowest LL with orbital quantum number N=0) but forms incompressible FQH states at $\nu=5/2, 7/2$ in the first excited LL (orbital quantum number N=1)\cite{willett_observation_1987}. In the MLG ZLL, orbital wave functions are identical to the N=0 LL of conventional semiconductor systems, and so no single-component even denominator FQH states are anticipated\cite{apalkov_fractional_2006, toke_theoretical_2007, abanin_fractional_2010, papic_tunable_2011, peterson_effects_2014}.
Multicomponent systems, however, can host a wider variety of FQH states\cite{halperin_theory_1983}, including at even denominator.  Indeed, EDFQH states at half \cite{suen_observation_1992, eisenstein_new_1992,liu_even-denominator_2014,liu_fractional_2014} and quarter\cite{luhman_observation_2008,shabani_evidence_2009} filling have been observed in the N=0 LL for structures where electrons are confined to two spatially separated layers or electronic subbands.

By analogy with such systems, it seems likely that the observed EDFQH states are multicomponent in nature, occurring at an underlying transition in the isospin order that is also responsible for the weakening of the ODFQH states.  In this scenario, the EDFQH states incorporate correlations between multiple nearly degenerate isospin components in the two phases.

\begin{figure}[ht!]
\begin{center}
\includegraphics[width=89mm, page=4]{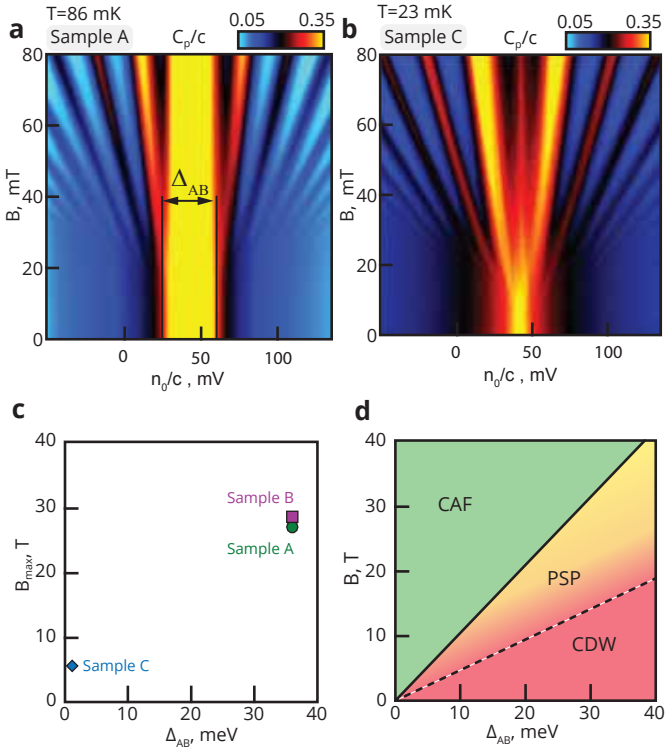}
\caption{
\textbf{
Low $B$ data and mean field phase diagram}
\textbf{(a)} Low field Landau fan in sample A, showing evidence of a large zero-field gap $\Delta_{AB}$ induced by sublattice splitting.
\textbf{(b)} Similar data for sample C, showing a much smaller sublattice gap.
\textbf{(c)} The estimated gap sizes for the three measured samples ($\Delta_{AB}$ plotted against the magnetic field at which the $\nu=\pm1/2$ states are maximal ($B_{max}$).
\textbf{(d)} Calculated phase diagram at $\nu=0$ for monolayer graphene as a function of magnetic field $B$ and sublattice splitting $\Delta_{AB}$.  The stability of the canted antiferromagnet (CAF), a sublattice polarized charge density wave (CDW), and a valley-coherent partially sublattice polarized phase (PSP) were calculated using anisotropy parameters $g_{\perp} = -10$ and $g_z = 15$ \cite{kharitonov_canted_2012} (see supplements).
\label{phasediagram}
}
\end{center}
\end{figure}

Unfortunately, it is difficult to experimentally determine the isospin order of the two phases.
In addition to the PSP and CAF phases, the mean field phase diagram of the charge neutral state in monolayer graphene\cite{kharitonov_phase_2012,kharitonov_canted_2012} includes a fully sublattice polarized charge density wave (CDW) state and fully spin polarized ferromagnetic (FM) state, with the latter distinguished by quantized helical edge states\cite{abanin_spin-filtered_2006,fertig_luttinger_2006}.
Previous experimental\cite{checkelsky_zero-energy_2008,young_tunable_2014} and numerical\cite{jung_theory_2009} work suggests that the charge neutral state at high fields is the CAF, which is favored by the anisotropic Coulomb interactions.
Large Zeeman energy favors the FM state, and an insulator-metal transition was observed using large in-plane magnetic fields\cite{young_tunable_2014}.  In devices rotationally aligned to hexagonal boron nitride substrates, the zero-$B$  band gap manifests in the ZLL as a large sublattice splitting that favors the CDW state. Experiments\cite{hunt_massive_2013,amet_insulating_2013} in gapped devices find a charge neutral state that is indeed gapped continuously from zero- to high-$B$ and is insensitive to in-plane magnetic fields, consistent with the CDW.

All the devices showing EDFQH states and the isospin transition share a similar  zero-field phenomenology, showing an insulating gapped state.  Figs.~\ref{phasediagram}a-b show low-$B$ Landau fan plots for samples A and C. The electron system remains incompressible, consistent with a single-particle $\Delta_{AB}$\cite{hunt_massive_2013,amet_insulating_2013}. The insulating nature of samples B and C is confirmed by transport measurements at zero field (Supplementary Fig.~S6; sample A did not have transport contacts). In a fourth sample showing no measurable sublattice gap, no EDFQH states were observed (Supplementary Figs.~S7-8). As shown in \ref{phasediagram}c, the magnetic field at which EDFQH states appear is directly correlated with $\Delta_{AB}$ as extracted from low-$B$ capacitance data (see also Figs.~S9).

The role of $\Delta_{AB}$ in determining the ZLL isospin order has been considered in \textit{bilayer} graphene \cite{kharitonov_phase_2012}, where sublattice splitting can be actuated with applied electric field, but is equally applicable in the present scenario. In addition to  $\Delta_{AB}$, the model  accounts for the spin Zeeman effect (with characteristic energy $E_Z=g\mu_B B\approx 1.34 K/T$) and valley anisotropic short range Coulomb interactions, characterized by energy $E_V\approx g_i\times\frac{a}{\ell_B}\frac{e^2}{\epsilon \ell_B}\approx g_i\times.98 K/T$ and parameterized by two dimensionless coupling constants, $g_z$ and $g_\perp$.  $g_z$ and $g_\perp$ are taken as $\Delta_{AB}$- and $B$-independent constants whose signs and magnitudes together favor either the CDW, CAF, FM, or PSP phase. The single-particle $E_Z$ favors the F phase, while $\Delta_{AB}$ favors the CDW phase.

Both $E_V$ and $E_Z$ grow with $B$ while $\Delta_{AB}$ is $B$ independent.  Thus the CDW phase is favored in the low-$B$ limit for $\Delta_{AB}\neq0$, with phase transitions to $E_V$- or $E_Z$- driven states possible at higher $B$. $g_z$ and $g_\perp$ are constrained to $g_\perp\approx -10$ and $g_z>-g_\perp$ from previous experiments\cite{checkelsky_zero-energy_2008,young_tunable_2014} (Supplementary information). Figure \ref{phasediagram}d shows the calculated phase diagram as a function of $\Delta_{AB}$ and $B$ for $g_\perp=-10$ and $g_z=15$. Two phase transitions are evident: a 2nd order transition from the CDW to PSP phase, corresponding to the canting of the valley order parameter into the plane, and a 1st order transition from the PSP to CAF phase. The critical $B$ for both transitions scales with $\Delta_{AB}$, matching the  experimental trend observed for the EDFQH states (Fig. \ref{phasediagram}c).  The absence of any measured in-plane magnetic field dependence (Figs.~S10-11) is also consistent with this model, in which the CAF state is nearly spin unpolarized due to the large value of $g_\perp$.  Finally, the $\nu$ dependence (Fig. 3e-f) is qualitatively consistent with the proposed mechanism: lattice-scale Coulomb anisotropies weaken with increasing $|\nu|$, so that a larger $B$ is required to reach the CAF phase.

\begin{figure}[hb!]
\begin{center}
\includegraphics[width=89mm, page=5]{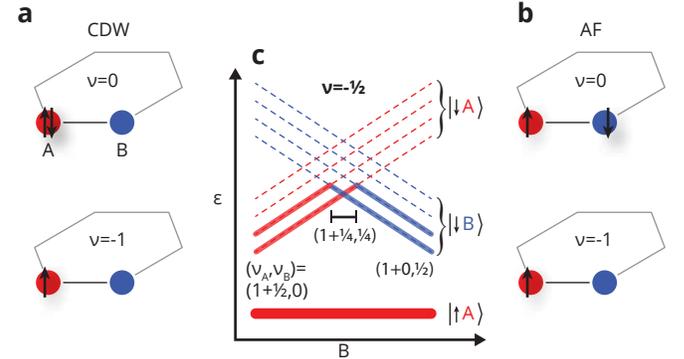}
\caption{
\textbf{Sublattice level crossing and multicomponent $\nu=\pm1/2$ states.}
\textbf{(a)} Spin and sublattice polarizations of $\nu=-1$ and $\nu=0$ integer quantum Hall ferromagnetic states in the low $B$ regime where the charge neutral ground state is in the CDW phase.
\textbf{(b)} Spin and sublattice polarizations of $\nu=-1$ and $\nu=0$ integer quantum Hall ferromagnetic states in the high $B$ regime, where the charge neutral ground state is in the AF phase.
\textbf{(c)} Level crossing at $\nu=-1/2$ (an identical scenario obtains at $\nu=+1/2$ by particle-hole conjugation across the ZLL). The $\nu=-1$ state is identical in both cases, consisting of a single fully sublattice- and spin-polarized LL, denoted $|\uparrow A\rangle$. On the CDW side of the transition, $\nu=-1/2$ consists of an additional half filled $|\downarrow A\rangle$ LL, while in the AF the half filled orbital is on the opposite sublattice, $|\downarrow B\rangle$; in both cases, half filling of a LL leads to compressible behavior.
When the two levels are nearly degenerate, Coulomb interactions favor a paired state with equal sublattice occupation $(\nu_A,\nu_B)=(1+1/4,1/4)$.   Bold lines indicate filled levels, the $|\downarrow A\rangle$ and $|\downarrow B\rangle$ levels are pictured as four  branches to depict partial occupation.
\label{levelcross}
}
\end{center}
\end{figure}

As the 2nd order CDW-PSP transition is crossed, the out-of-plane valley polarization of the CDW cants into the plane upon entry to the PSP phase. Charged single-particle excitations from these two states can be expected to be continuously related.  Absent distinct excitations at the transition, there is no basis for constructing multicomponent states. We thus hypothesize that EDFQH states and ODFQH transitions are associated with the PSP-CAF phase transition, which is first order.

The nature of the multicomponent states is transparent in the absence of both spin canting in the CAF phase and valley canting in the PSP phase. In this limit, the CDW-AF phase transition is direct, and increasing $B$ leads to a crossing between single-electron Landau levels with the same spin but localized on opposite sublattices (depicted schematically in Fig. 5). The $\nu=-1$ state is spin and sublattice polarized on both sides of the transition.  Additional electrons enter different sublattice orbitals in the two phases---in the CDW, electrons populate the same sublattice, while in the AF regime they populate the opposite sublattice. Far from the transition, then, single component FQH states are expected  as electrons fill the lower energy level. Half-filling an additional sublattice branch of lowest Landau level wavefunctions is closely analagous to the situation in single quantum wells, which results in a  compressible composite fermi liquid as observed in experiment.  Near the transition, however, intersublattice correlated states are possible as the energy cost of filling orbitals separated in energy is absent. The $-1/2$ state, for example, can be interpreted as a paired state of composite fermions\cite{halperin_theory_1983} on different sublattices. The $\nu=+1/2$ state is related by particle-hole conjugation across the full multicomponent Landau level. We note, however, that current theory does not provide a reason for why such a gapped state would be favored over a multicomponent composite Fermi liquid at the transition.

In addition, a number of experimental details do not fully conform with the level crossing picture.  This is perhaps unsurprising, as it explicitly neglects the subtle isospin polarizations of the CAF and PSP phases.
Recent experiments in graphene and GaAs quantum wells\cite{feldman_fractional_2013, maher_tunable_2014, liu_multicomponent_2015,zibrov_tunable_2017} have studied ODFQH at level crossings tuned by spin-, valley- or layer-Zeeman energy.  In all of these systems, ODFQH charge gaps repeatedly close and reopen, corresponding to a transfer of discrete fractional charge at each cyclotron guiding center between two or more isospin levels.  This leads to simple counting rules: for a FQH state built by populating $\nu=m/n$ electrons atop a given integer filling, $m+1$ distinct states are expected.

The observed phenomenology of the ODFQH states deviates from this mean-field composite fermion picture. In the two-flux sequence, ODFQH transitions do not feature full gap closing, but rather a weakening of charge gaps to a finite minimum (Fig.~3a) suggestive of an intermediate, interpolating  phase.  In sample C the $\pm3/5$ state does appear to close (Fig.~3b), although other ODFQH states show similar continuous crossovers between low- and high-$B$ phases.
Where gaps do close, most notably in the four-flux ODFQH states (Fig.~2), the simple counting rules are not followed, implying complex isospin physics. For example, in Fig.~2 three distinct gapped phases appear at $\nu = \pm 1/5$.
In a single-component Landau level, $1/5$ is the simplest composite fermion 4-flux state, corresponding to a single filled composite fermion level and admitting only full isospin polarization absent spontaneous coherence. The number of distinct $1/5$ states should thus directly count the number of distinct isospin orders at charge neutrality, implying an additional phase between PSP and CAF not predicted by the mean field model.   Alternatively, the $\pm1/5$ phases could be constructed from two component electron (hole) states built by adding $\pm4/5$ to the $\nu=\mp1$ state.  In this case, however, 5 distinct FQH states are expected rather than the three observed. Of course, the absence of similar phenomenology at $\nu = \pm 1/3$, the two-flux equivalent of the $\pm1/5$ states, also remain unexplained in the simple composite fermion picture.

Recent theoretical studies of the PSP-CAF transition have begun to address the shortcomings of mean field theory at the PSP-CAF transition.  Some authors have proposed that quantum fluctuations destroy the first order phase transition, leading to a deconfined critical point between the two phases\cite{lee_deconfined_2014,lee_wess-zumino-witten_2015}, while others suggest the first order phase transition survives but with an enlarged symmetry of low energy isospin rotations\cite{wu_so5_2014}.
In light of the present results, thermal transport measurements can be used to directly search for low-energy neutral excitations indicative of the predicted critical phases at $\nu=0$.
Others have addressed the intricate interplay of symmetry breaking and  fractional quantum Hall physics near this transition, which can lead to new filling-factor dependent isospin phases\cite{abanin_fractional_2013,sodemann_broken_2014}. However, none  have directly addressed the possibility of EDFQH states. We expect that future theoretical and experimental work---for example, measurements of tunneling exponents of EDFQH edge states---will be able to resolve the nature of these new phases.

\subsection*{Acknowledgements}
The authors acknowledge discussions with  D. Abanin, G. Murthy, Z. Papic, I. Sodemann,  and M. Zaletel and the experimental assistance of Scott Hannahs.  Magnetocapacitance measurements were funded by the NSF under DMR-1654186. A portion of the nanofabrication and transport measurements were funded by ARO under proposal 69188PHH. AFY acknowledges the support of the David and Lucile Packard Foundation. EMS acknowledges the support of the Elings Fellowship. The research reported here made use of shared facilities of the UCSB MRSEC (NSF DMR 1720256), a member of the Materials Research Facilities Network (www.mrfn.org).  Measurements above 14T were performed at the National High Magnetic Field Laboratory, which is supported by National Science Foundation Cooperative Agreement No. DMR-1157490 and the State of Florida.  K.W. and T.T. acknowledge support from the Elemental Strategy Initiative conducted by the MEXT, Japan and JSPS KAKENHI grant number JP15K21722.

\bibliographystyle{unsrt}

\pagebreak
\widetext
\begin{center}
\textbf{\large Supplemental Materials: Even denominator fractional quantum Hall states at an isospin transition in monolayer graphene}
\end{center}

\setcounter{equation}{0}
\setcounter{figure}{0}
\setcounter{table}{0}
\setcounter{page}{1}
\makeatletter
\renewcommand{\theequation}{S\arabic{equation}}
\renewcommand{\thefigure}{S\arabic{figure}}
\renewcommand{\bibnumfmt}[1]{[S#1]}
\renewcommand{\citenumfont}[1]{S#1}

\newcommand{\ev}{\epsilon_V}
\newcommand{\Dab}{\Delta_{AB}}
\newcommand{\ez}{\epsilon_Z}
\newcommand{\up}{u_\perp}
\newcommand{\uz}{u_z}
\newcommand{\gp}{g_\perp}
\newcommand{\gz}{g_z}

\newcommand{\sfigpath}{SuppFigsv4.pdf}
\def\lsim{\lower -0.3ex \hbox{$<$} \kern -0.75em \lower 0.7ex \hbox{$\sim$}}
\def\gsim{\lower -0.3ex \hbox{$>$} \kern -0.75em \lower 0.7ex \hbox{$\sim$}}
\def\mb{\bf}
\def\Vec#1{{\bf #1}}
\def\GVec#1{\mbox{\boldmath $#1$}}
\def\t{\theta}
\def\e{\epsilon}
\def\H{{\mathcal H}}
\def\vare{\varepsilon}
\def\av#1{\langle #1 \rangle}
\def\sgn{{\rm sgn}}
\def\partd#1#2{\frac{\partial #1}{\partial #2}}
\def\partdd#1#2{\frac{\partial^2 #1}{\partial #2^2}}
\def\kappaval{3}

\tableofcontents

\section{Materials and Methods}

We used the van der Waals dry transfer method to assemble graphite/hBN/MLG/hBN/graphite heterostructures. Graphite contact(s) were incorporated in the stack to contact the dual-gated monolayer. hBN thicknesses of 40-60 nm were used, while graphite contacts and gates were between 3 nm and 10 nm thick.  In samples A-C, windows to the graphite contacts and gates were etched in a Xetch-X3 xenon difluoride etching system, a selective hBN etch, and defluorinated with a 400$^{\circ}$C anneal in forming gas. The gates and contacts were then contacted Ti/Au (5nm/100nm) contacts. In sample D, edge contacts \cite{wang_one-dimensional_2013} to the graphite were made with Cr/Pd/Au (3nm/15nm/80nm). Optical images of the four measured devices are shown in Fig.~\ref{supp_sample_images}.


Measurements below $B$ = 14 T were performed in a top-loading Bluefors dry dilution refrigerator. Reported temperatures were measured using a ruthenium oxide thermometer attached to the cold finger. Higher magnetic field measurements were performed at the National High Magnetic Field Lab in Tallahassee in a 35 T resistive magnet and 45 T hybrid magnet, in He-3 fridges with a nominal base temperature of 0.3 K.
We performed measurements of the penetration field capacitance ($C_P$) as a function of magnetic field and gate voltages to probe incompressible/insulating states. This measurement technique is outlined in Fig.~\ref{supp_measurement} and described in detail in Ref.~\cite{zibrov_robust_2016} and references therein. Unless otherwise noted, measurements were performed \textit{above} the low frequency limit (at $f = 60$-$100$ kHz), i.e. there is an out of phase dissipative signal associated with many of the observed gapped states. In this frequency regime, an elevated $C_P$ indicates a combination of incompressibility and bulk insulating behavior, both are an indication of gapped states\cite{goodall_capacitance_1985}.  We focus on gapped states at fixed filling factor, which, by arguments
first proven by Streda\cite{streda_quantised_1982}, have quantized Hall conductance equal to their slope in the $n$-$B$ plane.

In. Fig.2a, a fixed filling factor running average of 3 pixels was used to remove line noise which obscured some weaker features. In Fig.3 c-d, a fixed filling factor running average of 5 pixels was used to remove line noise.

\section{Supplementary Figures}

\begin{figure*}[p!]
\begin{center}
\includegraphics[width=\columnwidth]{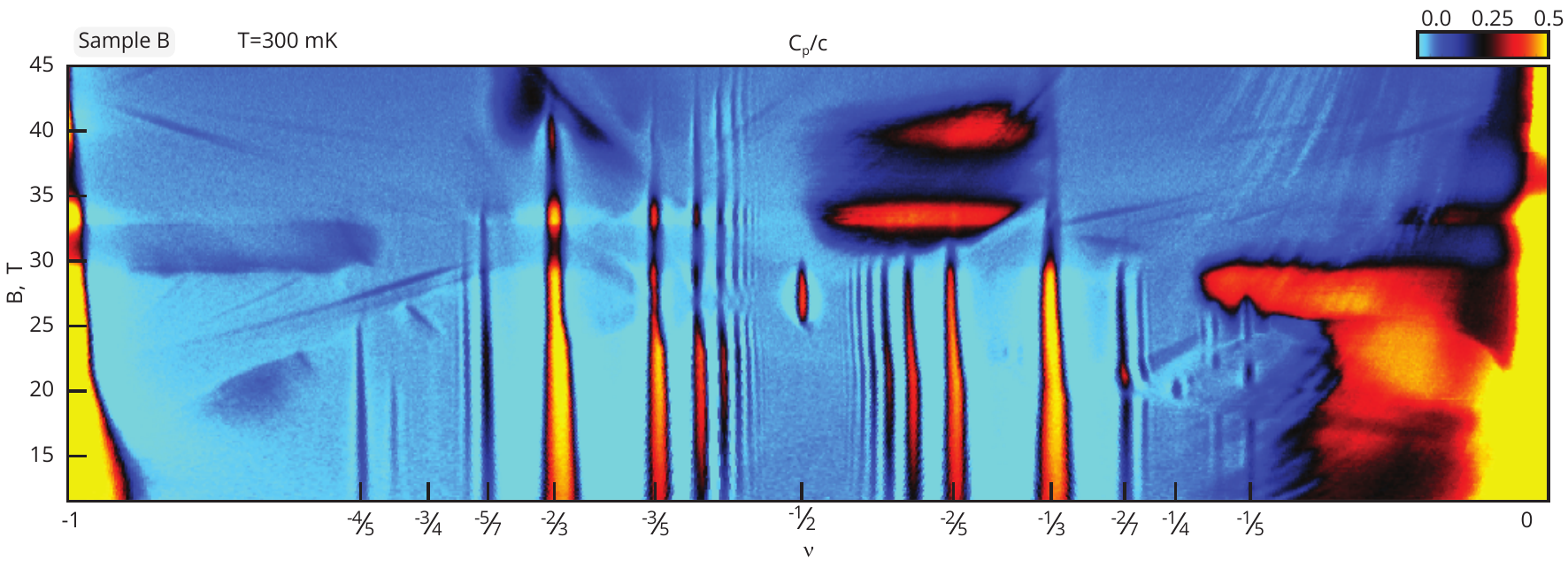}
\caption{\textbf{$C_P$ from $\nu = -1$ to $0$ in sample B.}
$C_P$ vs. filling factor and field, showing the relationship between the $\nu = -1/2$ and $\nu=-1/4$ states and their associated phase transitions. Above 30 T, all FQH states are suppressed by the features associated with the Hofstadter butterfly, which follow linear trajectories but not constant $\nu$.   Despite the similarity of the estimated zero-field gaps and magnetic field of $\nu = \pm 1/2$ states, Sample A shows no sign of a moir\'e pattern up to $B$ = 45T.
\label{supp_sampleB}
}
\end{center}
\end{figure*}


\begin{figure*}[p!]
\begin{center}
\includegraphics[width=\columnwidth]{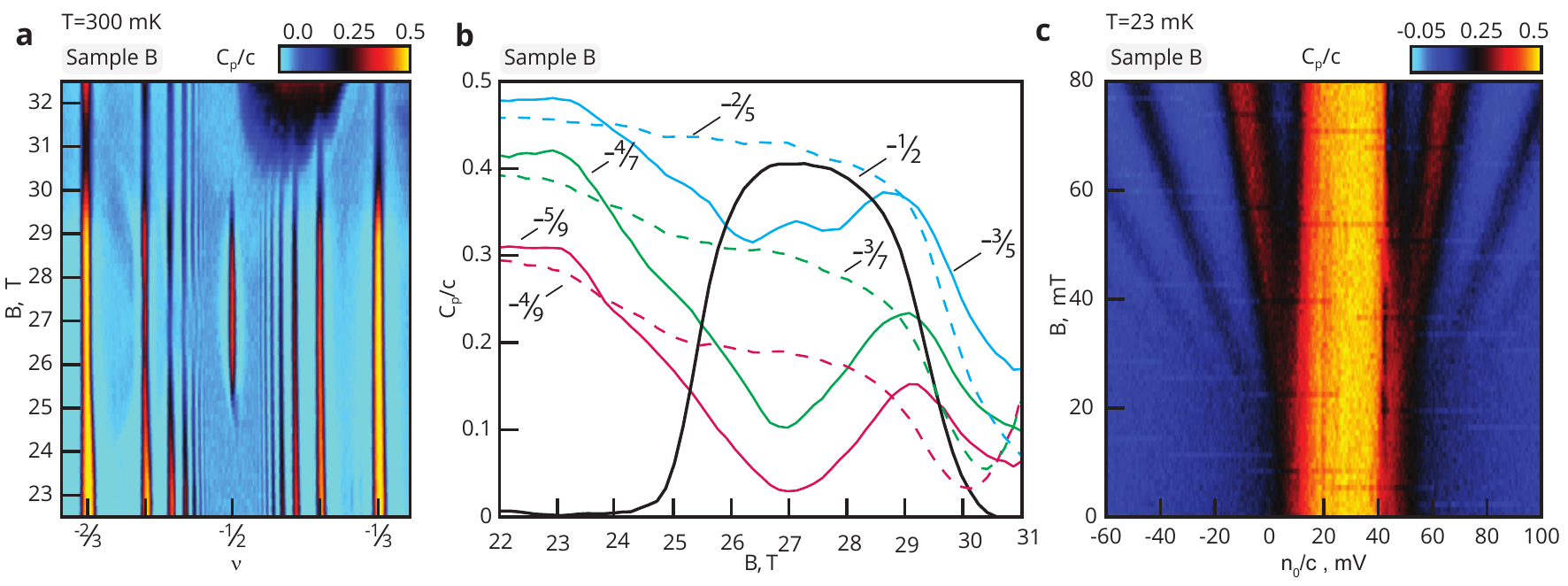}
\caption{ \textbf{Phenomenology of the $\nu = -1/2$ state in sample B.}
\textbf{(a)} $C_P$ as a function of $B$ and $\nu$ in the vicinity of $\nu=-1/2$ state for Sample B, taken at $T=300$ K
\textbf{(b)} $C_p$ peak height as a function of magnetic field plotted for selected FQH states $\nu\in(-1, 0)$ for Sample B.
\textbf{(c)} Low field Landau fan in Sample B, showing evidence of a large zero-field gap $\Delta_{AB}$ induced by sublattice splitting.
\label{supp_sampleB_gaps}
}
\end{center}
\end{figure*}


\begin{figure*}[p!]
\begin{center}
\includegraphics[width=\columnwidth]{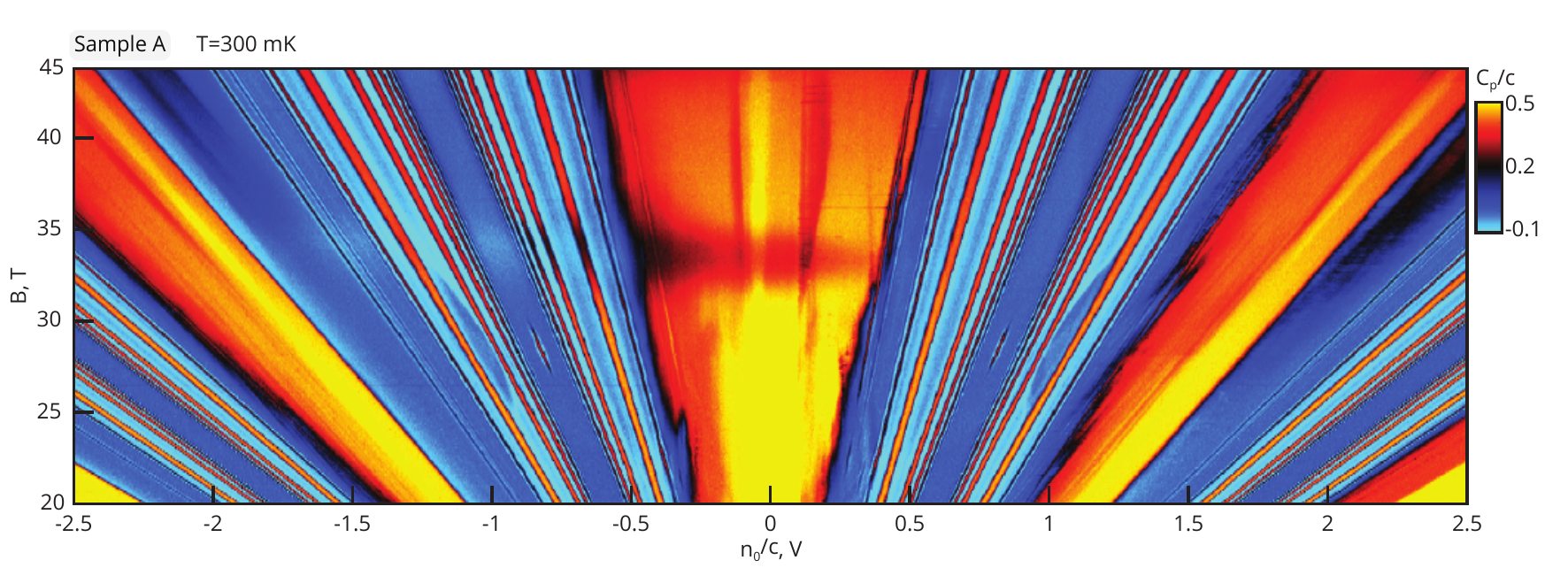}
\caption{\textbf{Expanded Landau Fan in sample A up to 45 T.}
 Penetration field capacitance ($C_P$) as a function of charge carrier density ($n_0/c$) and magnetic field ($B$) in sample A.
\label{supp_sampleA_LF}
}
\end{center}
\end{figure*}


\begin{figure*}[p!]
\begin{center}
\includegraphics[width=\columnwidth]{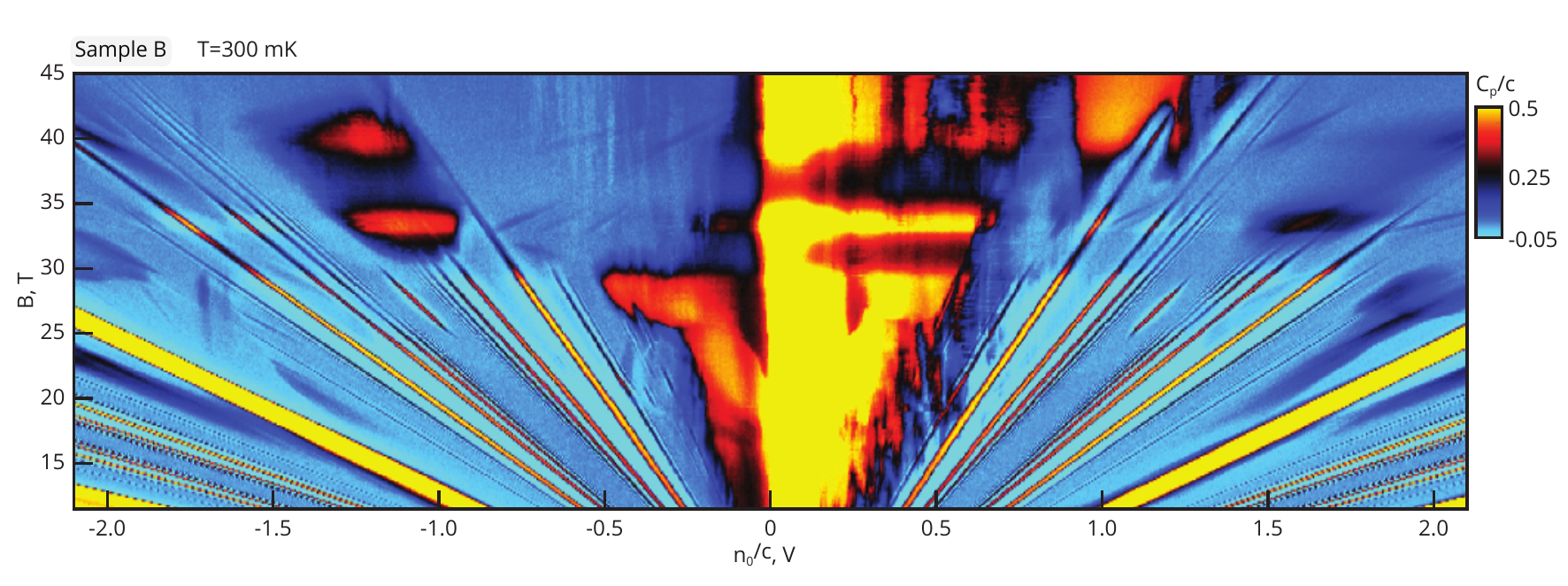}
\caption{\textbf{Expanded Landau Fan in sample B up to 45 T.}
Penetration field capacitance ($C_P$) as a function of charge carrier density ($n_0/c$) and magnetic field ($B$) in sample B. Despite similarities in the estimated zero field gap ($\Delta$) between sample A and sample B, sample B exhibits a weakening of the fractions and Hofstadter features with a full flux quantum per unit cell at $B=43 T$, while sample A does not show any strong Hofstadter features.
\label{supp_sampleB_LF}
}
\end{center}
\end{figure*}


\begin{figure*}[p!]
\begin{center}
\includegraphics[width=\columnwidth]{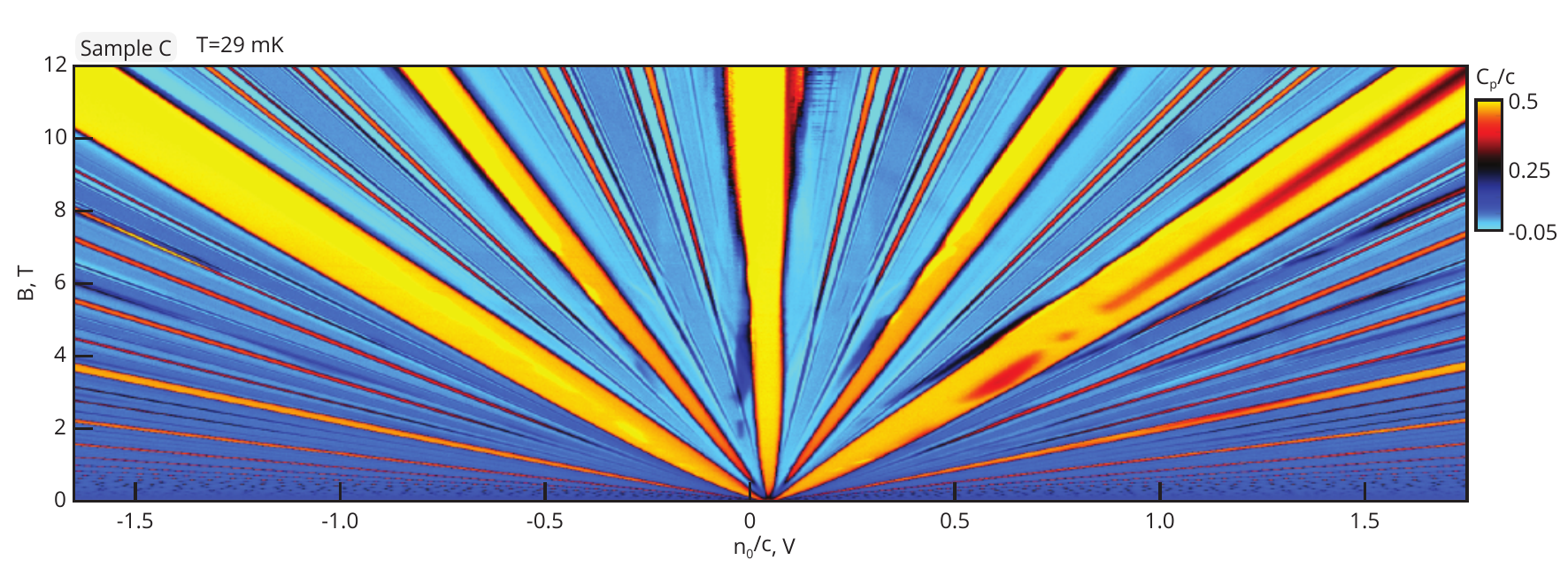}
\caption{\textbf{Expanded Landau Fan in sample C up to 12 T.}
Penetration field capacitance ($C_P$) as a function of charge carrier density ($n_0/c$) and magnetic field ($B$) in sample C. The four flux FQH states are not well developed by $B$ = 6 T, most likely preventing the observation of $\nu = \pm 1/4$ states in this device.
\label{supp_sampleC_LF}
}
\end{center}
\end{figure*}


\begin{figure*}[p!]
\begin{center}
\includegraphics[width=\columnwidth]{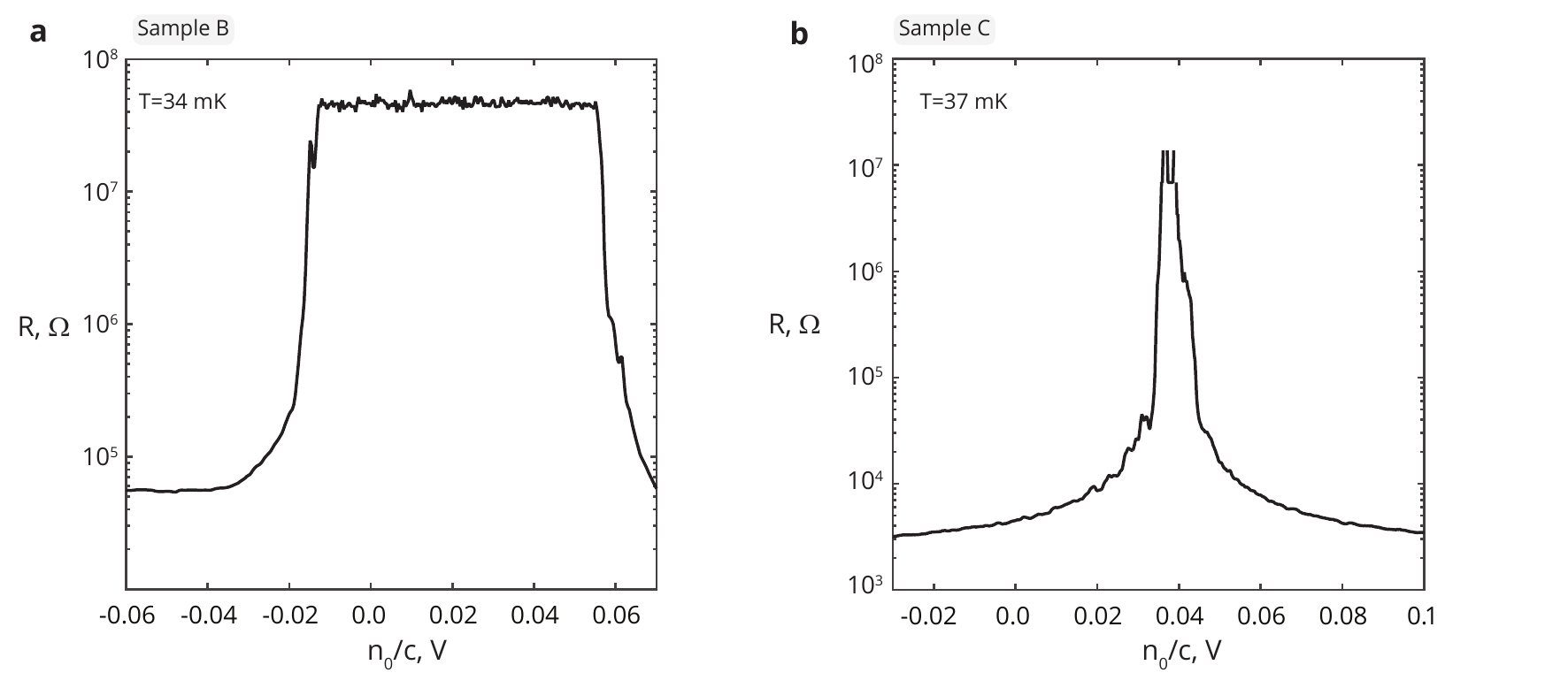}
\caption{ \textbf{Two terminal transport in sample B and C.}
Samples were voltage biased with an RMS amplitude of 100 $\mu V$ and the current was measured.
\textbf{(a)} Two-terminal resistance as a function of $n_0$ in sample B at T = 34 mK.  The high resistance regime is cut off by the input impedance of our lock-in amplifier.
\textbf{(b)} Two-terminal resistance as a function of $n_0$ in sample C at T = 37 mK, showing a much narrower insulating regime than sample B.
}
\label{supp_transport}
\end{center}
\end{figure*}


\begin{figure*}[p!]
\begin{center}
\includegraphics[width=\columnwidth]{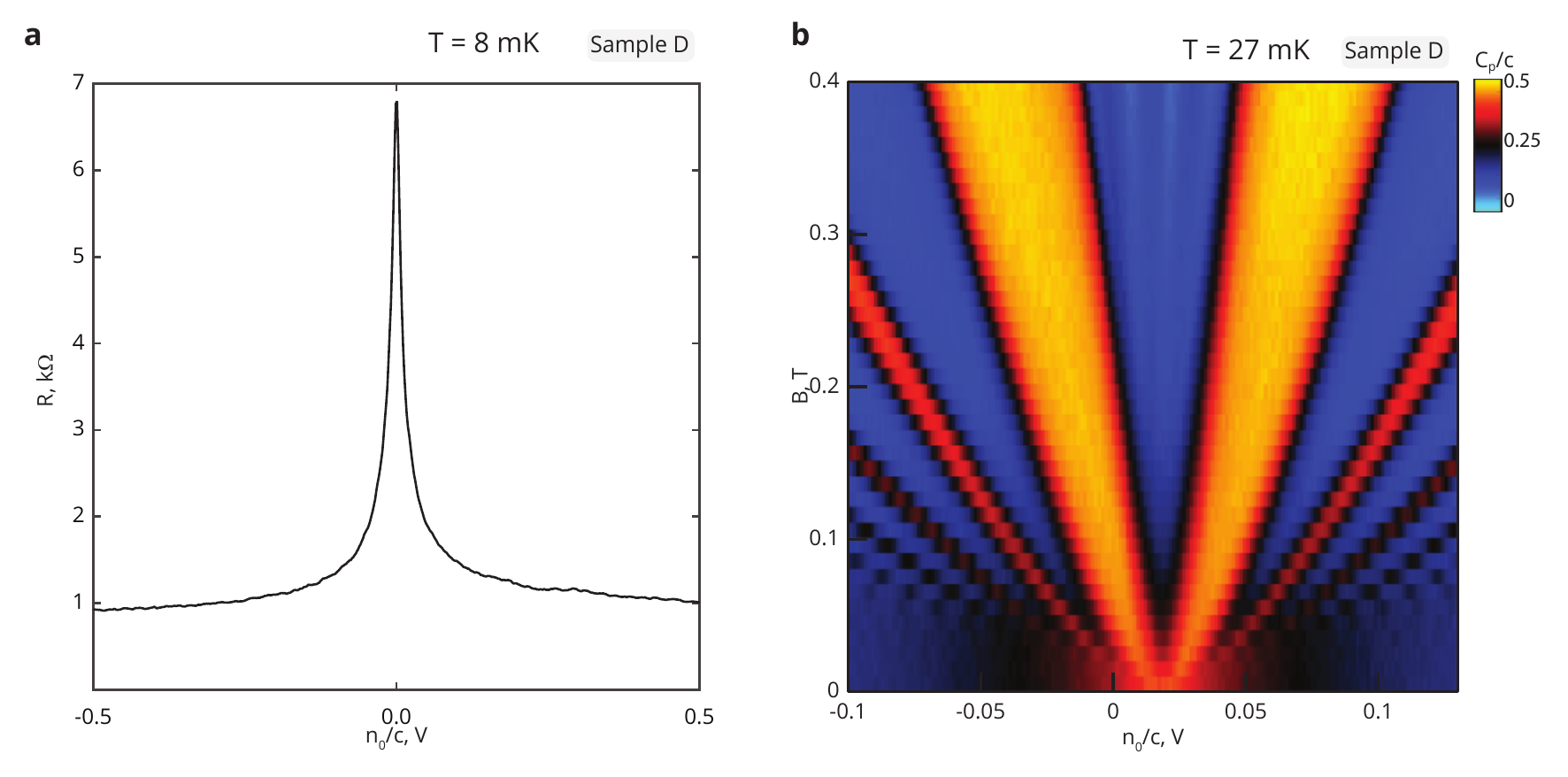}
\caption{\textbf{Evidence of the absence of a sublattice gap in sample D.}
\textbf{(a)} Two terminal transport in sample D at zero applied magnetic field. The device was voltage biased at 100 $\mu V$ and the induced current was measured. \textbf{(b)} Low field $C_P$ Landau fan in sample D, showing the absence of an incompressible peak at  $\nu=0$ just above $B=0$, in contrast with the behavior of gapped samples (Fig.~4a,b).
\label{supp_sampleD_lowfield}
}
\end{center}
\end{figure*}

%
\begin{figure*}[p!]
\begin{center}
\includegraphics[width=\columnwidth]{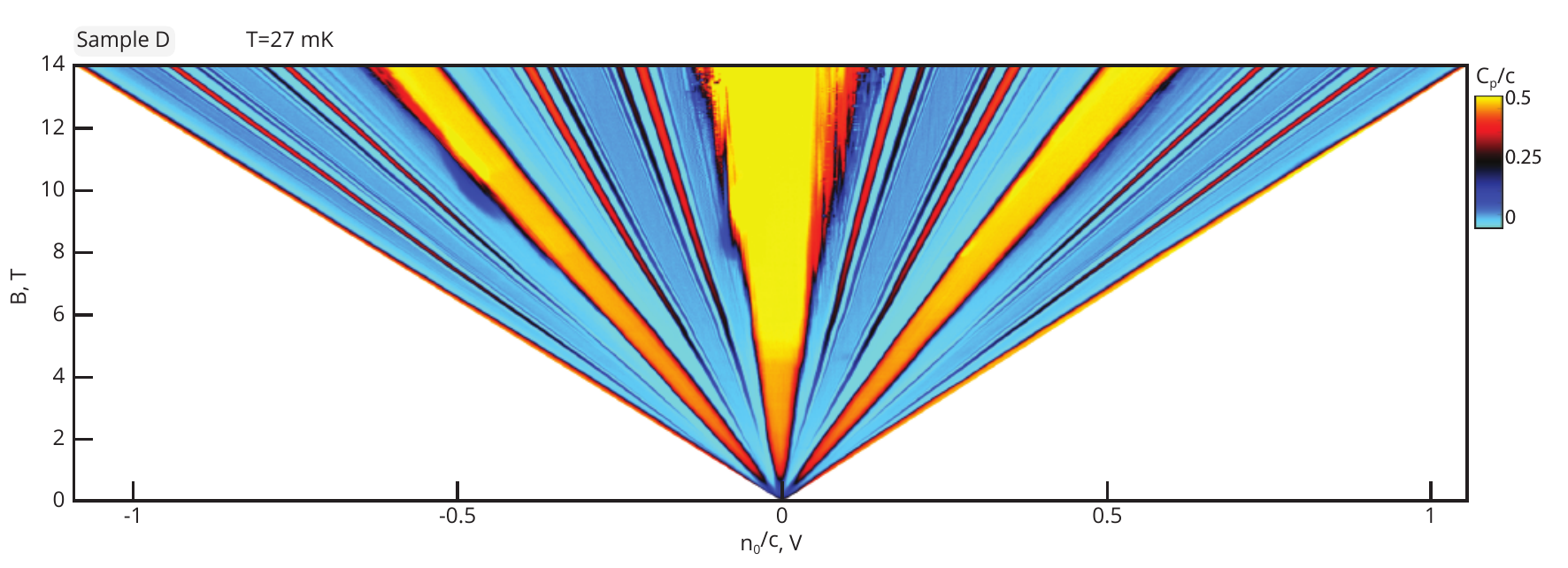}
\caption{\textbf{$C_P$ Landau fan in sample D.}
$C_P$ as a function of gate voltage ($V_G$) and applied magnetic field ($B$). This monolayer sample shows no zero-field insulating gap, and exhibits phase transitions in FQH states in $\nu\in[-2,-1]$ and $\nu\in[1,2]$ which are not observed in any of the zero-field gapped samples. $\nu = \pm1/2$ states are not observed up to 14 T.
\label{supp_sampleD_LF}
}
\end{center}
\end{figure*}

%
\begin{figure*}[p!]
\begin{center}
\includegraphics[width = \columnwidth]{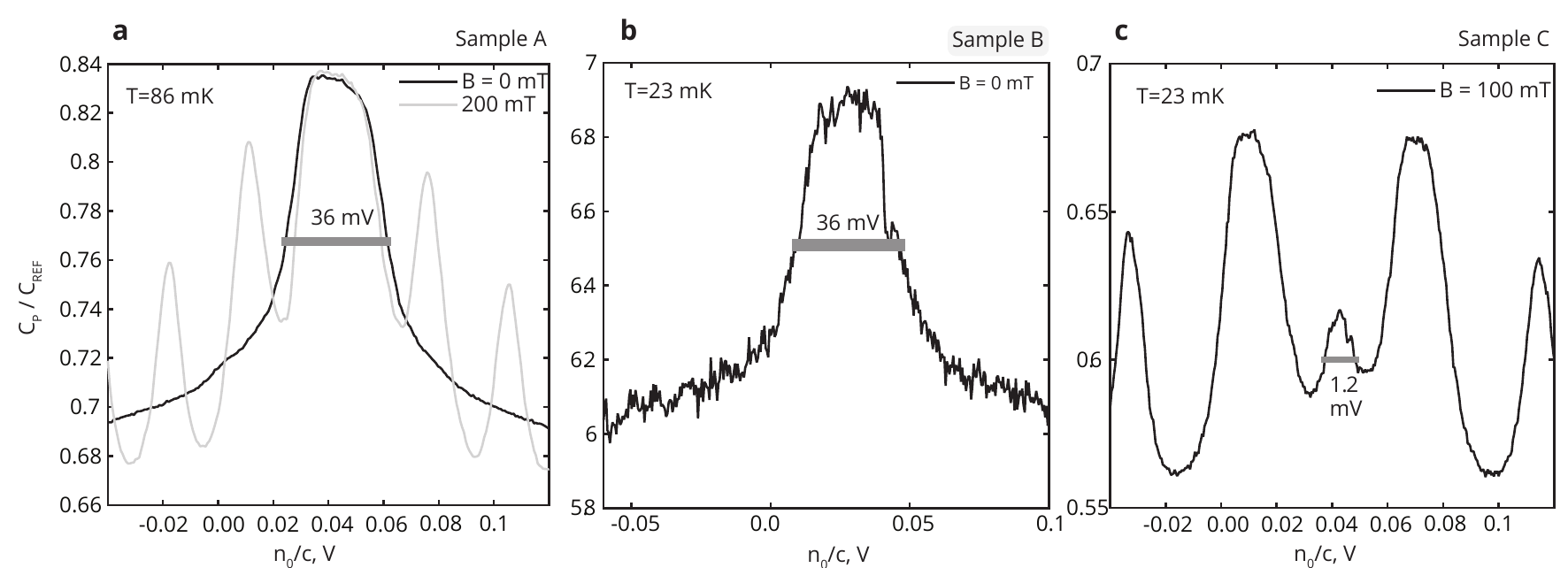}
\caption{ \textbf{Estimation of zero-field gaps ($\Delta$) from capacitance.}
\textbf{(a)} Line cuts of $C_P/C_{REF}$, where $C_{REF}$ is an extern reference capacitor, as a function of $n_0$ at B=0 mT (black) and 200 mT (gray) in sample A extracted from Fig.~2e in the main text. The gap is estimated to be $\Delta = 36 mV$ by the observed width of the peak in $C_P$, which is mainly determined by the quantum capacitance.
\textbf{(b)}  Line cut of $C_P$ at $B$ = 100 mT for sample B,  extracted from Fig.~4a in the main text. Here, the estimation of the gap is made difficult by the presence of density of states effects at zero field and the presence of LLs at finite field. An estimate of the gap is shown in gray, and a similar scale of gap is estimated from transport (see Fig.~S6).
\textbf{(c)} Line cut at B=0 mT in sample C extracted from Fig.~4b of the main text. The gap is estimated to be $\Delta = 36 mV$, similar to sample A.
}
\label{supp_gaps}
\end{center}
\end{figure*}


\begin{figure*}[p!]
\begin{center}
\includegraphics[width=\columnwidth]{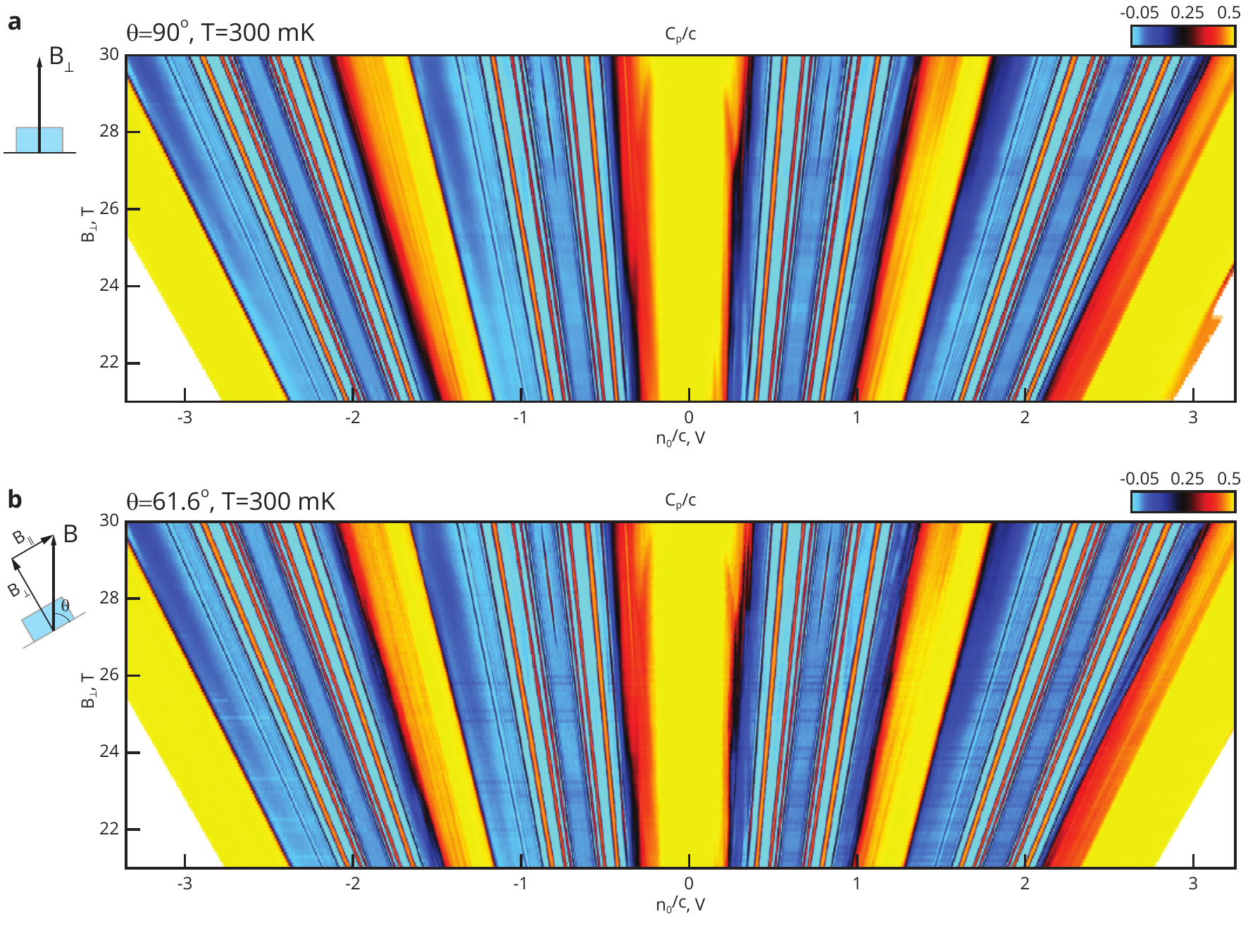}
\caption{\textbf{Dependence of $\nu = \pm1/2$ on tilted magnetic field in Sample A.}
$C_P$ as a function of density ($n_0/c$) and applied perpendicular magnetic field ($B$) with \textbf{(a)} the sample fully perpendicular to the field ($\theta = 90^{\circ}$) and \textbf{(b)} tilted to $\theta = 61.6^{\circ}$. The $\nu = \pm 1/2$ FQH states do not change with an applied in-plane magnetic field.
\label{supp_sampleA_tilt}
}
\end{center}
\end{figure*}


\begin{figure*}[p!]
\begin{center}
\includegraphics[width=\columnwidth]{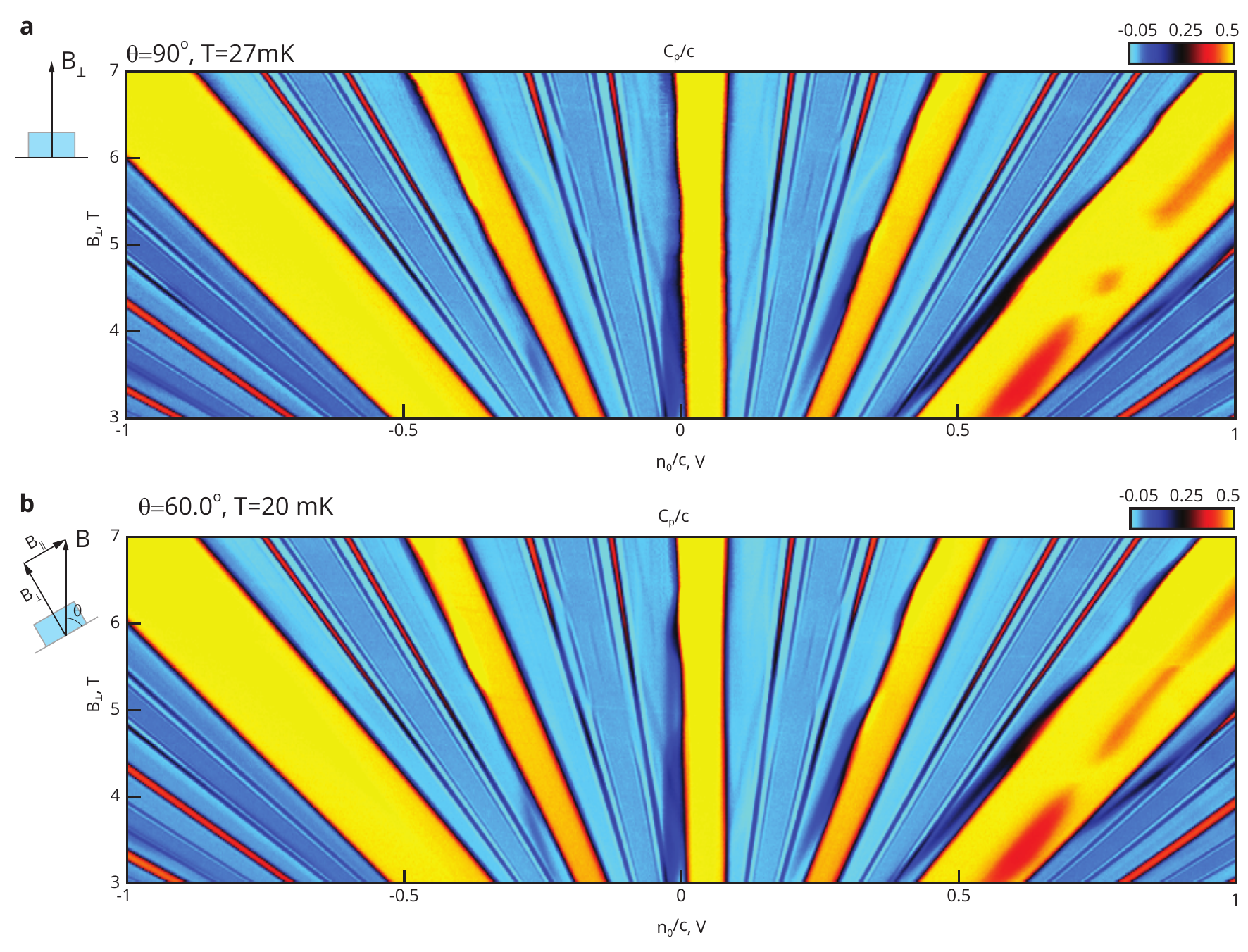}
\caption{\textbf{Dependence of $\nu = \pm1/2$ and FQH transitions on tilted magnetic field in Sample C.}
$C_P$ as a function of density ($n_0/c$) and applied perpendicular magnetic field ($B$) with \textbf{(a)} the sample fully perpendicular to the field ($\theta = 90^{\circ}$) and \textbf{(b)} tilted to $\theta = 60.0^{\circ}$. The $\nu = \pm 1/2$ FQH states and the transitions in odd denominator FQH states do not change with an applied in plane magnetic field, suggesting that spin polarization does not play a significant role in the phase transition associated with both phenomena.
\label{supp_sampleC_tilt}
}
\end{center}
\end{figure*}

\begin{figure*}[p!]
\begin{center}
\includegraphics[width=90mm]{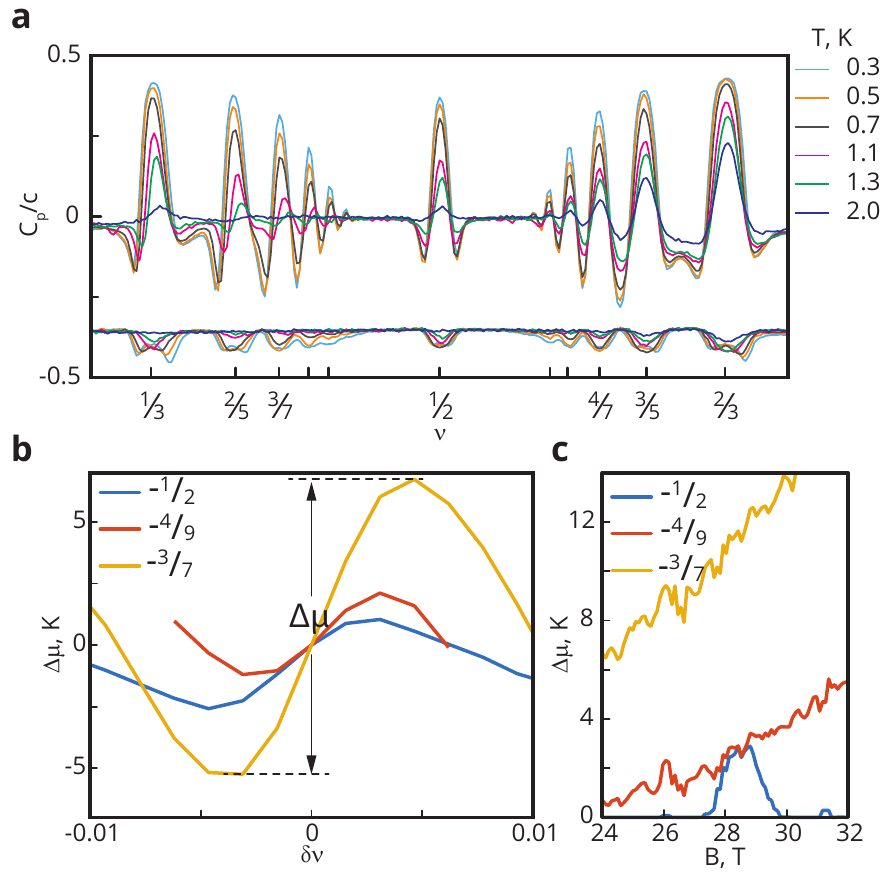}
\caption{\textbf{Temperature dependence and thermodynamic gap measurements in sample A.}
\textbf{(a)} Penetration field capacitance $C_\mathrm{p}/c$ (upper curves) and
 dissipation(lower curves) at $B=28.3$ T, $\nu=[1/3, 2/3]$ as a function of
 temperature (measured in K). The incompressible state at $\nu=1/2$ is still
 present at $T=2$ K.
\textbf{(b)} Thermodynamic gap $\Delta\mu \sim \frac{e}{k_B} \int \frac{C_\mathrm{p}}{c}d(n_0/c)$
at $T=1.6 K$. The comparatively high temperature is chosen to ensure sufficient conductivity to reach the low-frequency regime\cite{goodall_capacitance_1985}: as $T\to0$ the gapped bulk becomes exponentially insulating and the measured penetration field capacitance is no longer an accurate probe of density of states (see Methods).
\textbf{(c)} Thermodynamic gap as a function of magnetic field for $\nu=-1/2$, $\nu=-3/7$ and $\nu=-4/9$ at T=1.6 K.
At its largest $\Delta\mu_{-1/2}=2.8\mathrm{K}\approx\Delta\mu_{-4/9}$.
\label{supp_temp_dep}
}
\end{center}
\end{figure*}


\begin{figure*}[p!]
\begin{center}
\includegraphics[width=90mm]{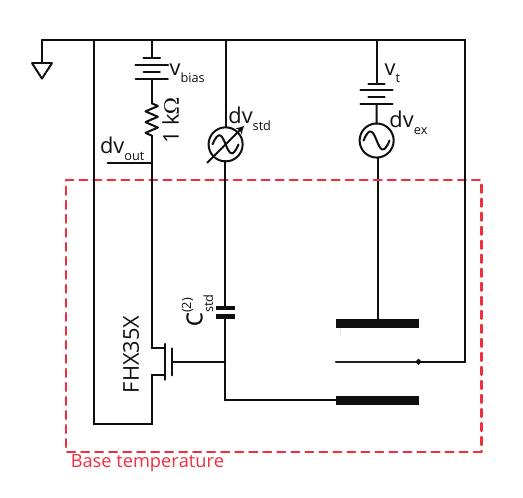}
\caption{\textbf{Measurement schematic.}
The penetration field capacitance $C_P$ is measured in a capacitance bridge configuration against a fixed, on-chip reference capacitor $C_{std}$. A fixed AC excitation is applied to the sample ($dv_{exc}$) and a variable phase and amplitude ac excitation of the same frequency ($dv_{std}$) is applied to the reference capacitor to balance the capacitance bridge. The voltage at the balance point is amplified by a low temperature transistor (FHX35X) which is biased with a 1 $k\Omega$ resistor. In the measurements presented here, the bridge is balanced at the beginning of a measurement at a fixed location and deviations from balance are measured as the dc voltages of the sample and/or gate are swept.
\label{supp_measurement}
}
\end{center}
\end{figure*}


\begin{figure*}[p!]
\begin{center}
\includegraphics[width=170mm]{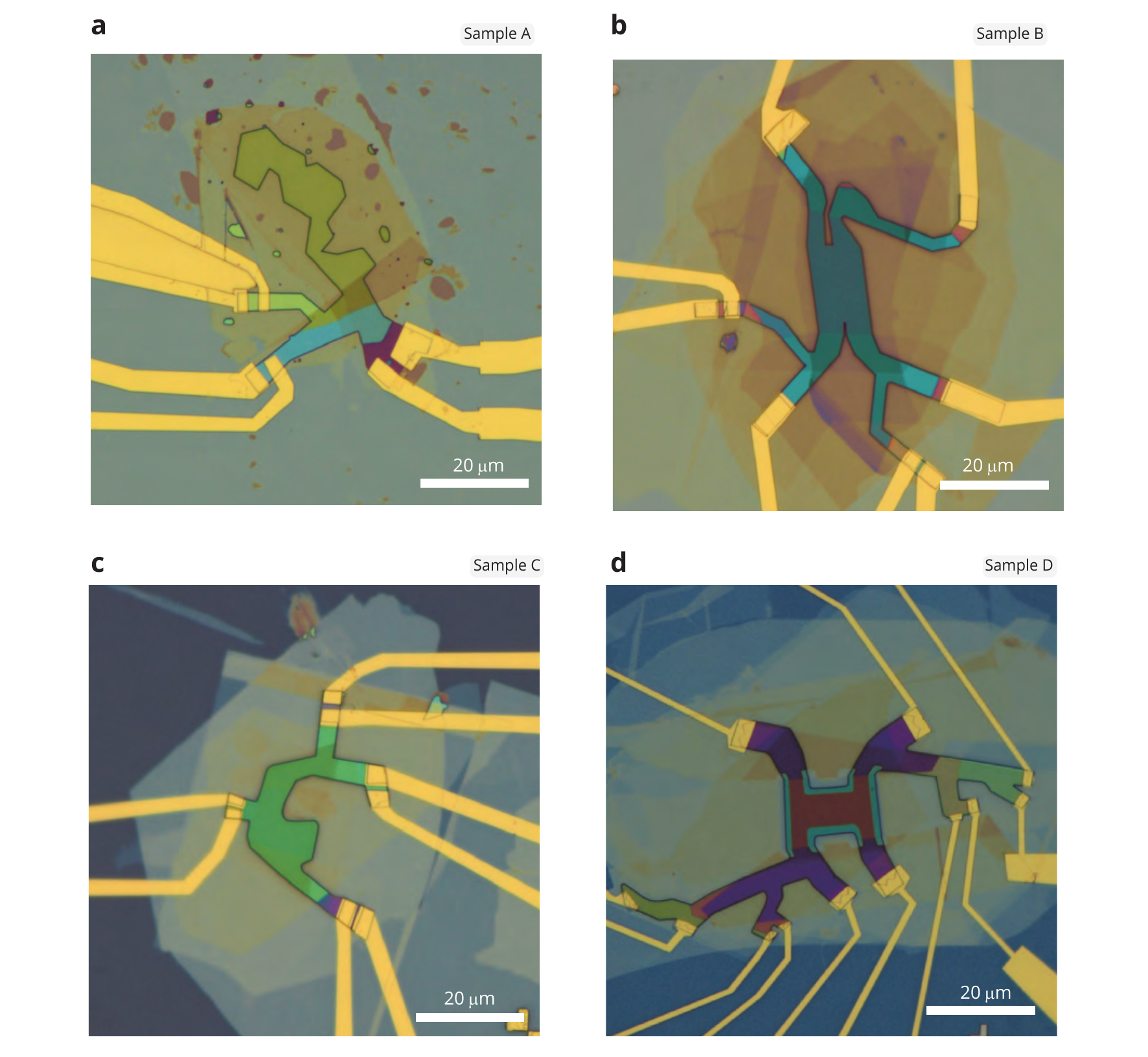}
\caption{\textbf{Optical images of the measured samples.}
\textbf{a-d.} Optical images of samples A-D, respectively. Scale bar is 20 $\mu m$.
\label{supp_sample_images}
}
\end{center}
\end{figure*}

\clearpage
\section{Calculation of $\nu=0$ phase diagram}
Here we describe the calculation of the $\nu=0$ phase diagram, based on the model outlined in \cite{kharitonov_canted_2012}.  This model, originally intended for bilayer graphene, is in fact a generalization of a monolayer model\cite{kharitonov_phase_2012} to include a
valley Zeeman that neglects bilayer physics, such as the orbital degeneracy, that would make it inapplicable to monolayer. Within this model, there are four phases: a spin polarized, valley-singlet Ferromagnet (F), a valley polarized, spin-singlet lattice-scale charge density wave (CDW), a canted antiferromagnet that is partially spin polarized and valley unpolarized (CAF), and a spin-singlet partially sublattice polarized phase (PSP), in which the valley polarization lies somewhere between the z-axis and the xy-plane. In the limit of $\Delta_{AB}=0$, the PSP phase becomes full sublattice unpolarized (with valley polarization lying in the plane). This phase is known as the Kekule distorted phase (KD) in the literature, and in this limit, the KD-CDW phase transition is also first order.  Following the terminology of reference \cite{kharitonov_canted_2012}, the PSP phase is analogous to the partially layer polarized phase (PLP) while the CDW phase is analogous to the Fully Layer Polarized (FLP) phase.

The energies of the different phases, and their phase boundaries, are obtained by calculating the energy expressions for explicit forms of the isospin wavefunctions in the expressions provided in \cite{kharitonov_canted_2012}. They are:
\begin{equation}
  \begin{array}{|c|c|c|}
\hline
\textrm{phase}&\textrm{energy}& \textrm{condition}\\  \hline
\textrm{F}&-2 \ez-2 \up-\uz&\\ \hline
\textrm{CDW}&\uz-2 \Dab&\\ \hline
\textrm{CAF}&   -\uz-\frac{\ez^2}{2|\up|}&0 < -\frac{\ez}{2\up} < 1\\ \hline
\textrm{PSP}&\up-\frac{\Dab^2}{\uz+|\up|}&\frac{\Dab}{\uz + |\up|} < 1\\\hline
\end{array}
\label{anisotropyenergies}
\end{equation}

Here $\Dab$ is the single particle AB sublattice splitting; $\ez$ is the Zeeman energy, and $u_{z,\perp}=g_{z,\perp}\frac{a}{\ell_B}\frac{2^2}{\epsilon \ell_B}$ are the anisotropic interaction energies as described in the main text.

\begin{figure*}[ht!]
\begin{center}
\includegraphics[width = \columnwidth]{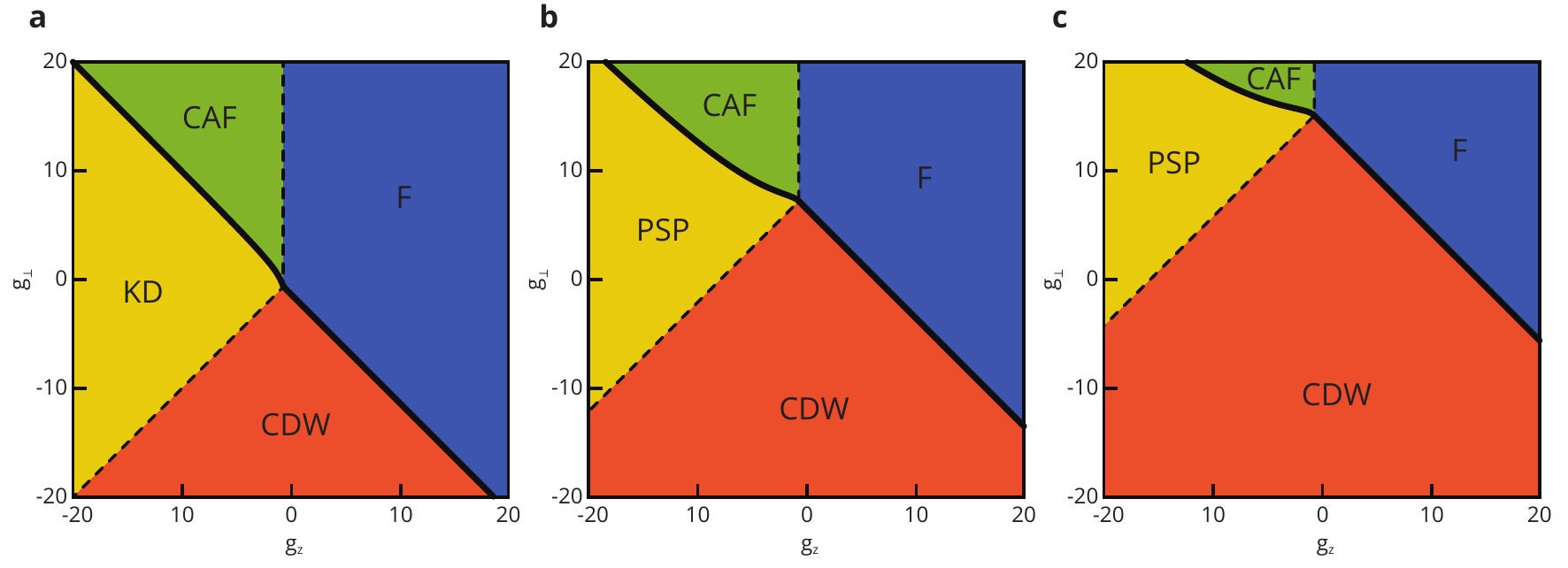}
\caption{\textbf{Phase diagram for different values of $\Delta_{AB}$.}  \textbf{(a)} For $\Delta_{AB}=0$, the phase diagram is magnetic field independent and the CDW-KD transition is first order.  \textbf{(b)} Finite $\Dab=10$ meV stabilizes the CDW phase while transforming the CDW-KD phase transition to 2nd order. \textbf{(c)} $\Dab=20$ meV.
\label{gpgz}
}
\end{center}
\end{figure*}

The energies depend on four parameters: $\gz$, $\gp$, $\ez$, and $\Dab$.  $\ez=g\mu_B B_T$ follows from the fact that spin-orbit coupling is exceptionally weak in graphene, so that $g=2$.  $\Dab$ we estimate from the low field behavior in each device, giving $\sim$25mV for device A and $\sim$2 meV for device B, and $\sim$0mV for device C. Phase diagrams in the $\gp-\gz$ plane for different values of $\Dab$ are shown in Fig. \ref{gpgz}.

The interaction anisotropy parameters are more difficult to estimate, and follow from knowledge of the $\nu=0$ phase diagram.
For example the phase transition from the CAF to the F occurs at $u_p=-\ez/2$, and can thus be tuned by varying the total- and in-plane magnetic fields independently. Notably, these are determined at high energies, and are unlikely to be affected significantly by low energy band structure effects such as the presence of a $\Dab$ gap, allowing us to estimate them from experiments on $\Dab=0$ samples. Reference \cite{young_tunable_2014} reported this crossover at out-of-plane magnetic fields of $B_\perp\approx 1T$ and total magnetic field $B_T\approx 25T$ in single-gated graphene devices.  From this measurement, we estimate
\begin{align}
\up&=\frac{\ez}{2}\\
\gp \frac{a}{\ell_{B_\perp}}\frac{e^2}{\frac{\epsilon_{hBN}+\epsilon_{vac}}{2} \ell_{B_\perp}}&=-\frac{g\mu_B B_T}{2} \\
\gp&\approx-10\label{gp}
\end{align}
where we use the in-plane dielectric constant of hBN, $\epsilon_{hBN}=6.6$, most relevant for screening of Coulomb interactions.

This leaves $g_z$ as a free parameter.  We assume that gapless ($\Dab=0$), neutral graphene at high field is in the CAF state.  The evidence for this is somewhat circumstantial: based solely on \cite{young_tunable_2014}, the KD phase cannot be excluded.  However, as will be evident below, if this is not the case then no phase transition to the CAF in the current experiment would be possible, inconsistent with our
observation of first-order phase transitions in the FQH regime.  From Eq. \ref{anisotropyenergies}, we can derive the following constraints:
\begin{widetext}
  \begin{align}
  \mathcal{E}_{CAF}&<\mathcal{E}_{F}&\mathcal{E}_{CAF}<\mathcal{E}_{KD}\\
  -\uz-\frac{\ez^2}{2|\up|}&<-2\ez-2\up-\uz&-\uz-\frac{\ez^2}{2|\up|}&<\up\\
  \up&<-\frac{1+\sqrt{2}}{2}\ez&            \uz&>\frac{\ez^2-2\up^2}{2\up} \\
  \gp&\lesssim -1.7&\gz&\gtrsim -\gp+\frac{.93}{\gp}
\end{align}
\end{widetext}

\begin{figure*}[ht!]
\begin{center}
\includegraphics[width=\columnwidth]{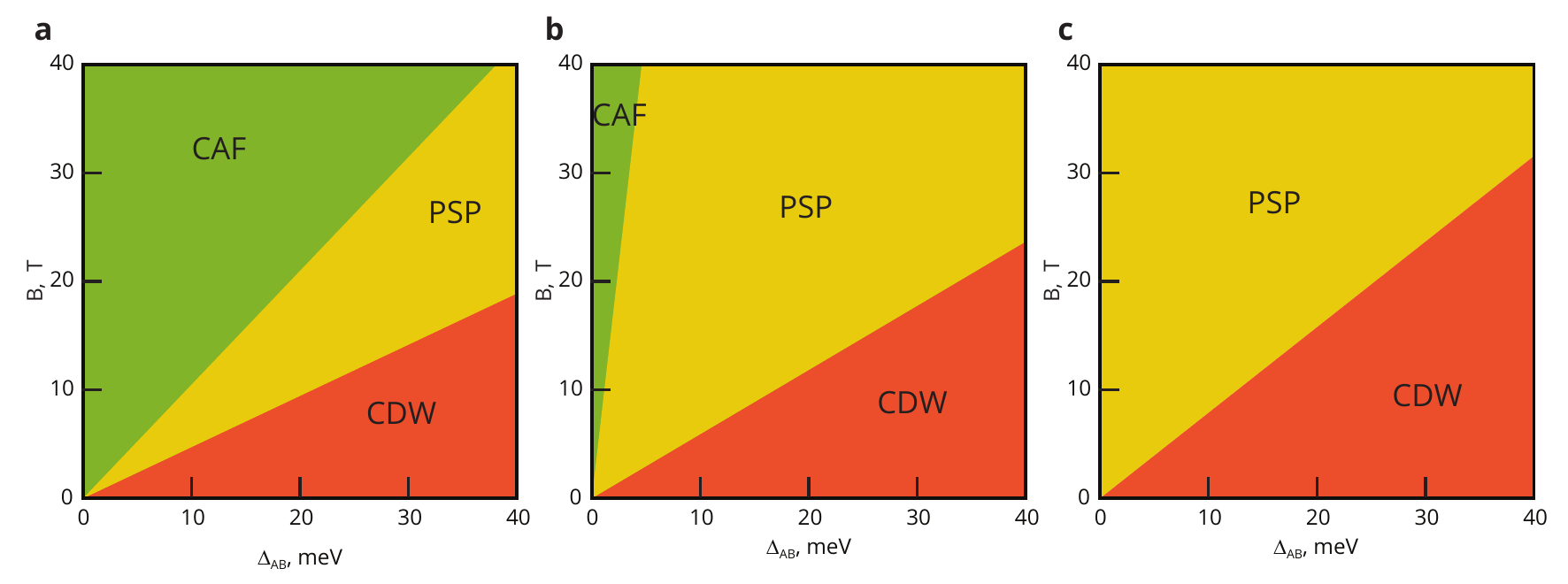}
\caption{\textbf{Phase diagram in the $B$-$\Delta_{AB}$ plane.}  Theoretically calculated phase diagram for (a) $\gz$=15, (b) $\gz$=10, and (c) $\gz$=5.
\label{DabB}
}
\end{center}
\end{figure*}

As described in the main text, with the exception of $\Dab$, all the anisotropy energies are linear in $B$.
Thus the low $B$ limit is equivalent to very large $\Dab$, where we expect the CDW ground state to prevail.  Conversely, the high $B$ limit in our device is equivalent to the low $\Dab$ limit in \cite{kharitonov_canted_2012}. Figure \ref{DabB} shows the phase diagram for different values of $g_z$.  For choices of $\gz$ such that the `natural' state is indeed the CAF, this state will obtain at the highest magnetic field, with the magnetic field required tuned by $g_z$ as well as $\Delta_{AB}$.

\end{document}